\documentstyle[11pt,aaspp,psfig]{article}
\begin{document}

\title{HST Observations of the Gravitationally Lensed Cloverleaf 
Broad Absorption Line QSO H1413+1143: 
Modeling the Lens\footnote{Based on observations with the NASA/ESA 
{\it Hubble Space Telescope}, obtained at the Space Telescope Science
Institute, which is operated by AURA, Inc., under NASA contract NAS 5-26555.}}
\author{Kyu-Hyun Chae and David A. Turnshek}
\affil{Department of Physics \& Astronomy, University of Pittsburgh,
Pittsburgh, PA 15260}

\begin{abstract}
We investigate gravitational lens models for the quadruply-lensed
Cloverleaf BAL QSO H1413+1143 based on the HST WFPC/WFPC2 
astrometric and photometric data of the system by Turnshek et al.\
and the HST NICMOS-2 data by Falco et al. 
The accurate image positions and the dust-extinction-corrected relative
amplifications, along with a possible detection of the lensing galaxy in the
infrared, permit more accurate lens models than were previously
possible. While more recent models are qualitatively 
consistent with the HST data, none of the previous models considered 
the dust-extinction-corrected relative amplifications of the image 
components.  We use the power-law elliptical mass model to fit the HST 
data. We find that a single elliptical galaxy perturbed by an external 
shear can fit the image positions within the observational uncertainties; 
however, the predicted relative magnifications are only roughly 
consistent with the observational relative amplifications.
We find that a primary galaxy combined with a secondary galaxy in the 
vicinity of the Cloverleaf or a cluster centered (south-)west 
of the Cloverleaf can fit both the image positions and relative 
amplifications within the observational uncertainties. 
We discuss future observations which could be used to test and/or 
further constrain lens models of the Cloverleaf.
\end{abstract}


\section{Introduction}

The gravitationally-lensed ``Cloverleaf'' consists of four bright
image components of a distant Broad Absorption Line (BAL) QSO
H1413+1143 ($z \approx 2.55$;  $V \approx 16.8$), which was
discovered by Hazard et al.\ (1984) and later identified to be
a lens system by Magain et al.\ (1988). Since its discovery
the search for the lens in optical wavebands has not resulted in
a detection (e.g.\ Turnshek et al.\ 1997; hereafter T97) until
just recently (Falco et al.\ 1998). A recent analysis of the 
Falco et al.\ HST NICMOS-2 data has shown some evidence for the
lensing galaxy near the geometrical center of the Cloverleaf 
(Kneib, Alloin, \& Pell$\acute{\mbox o}$ 1998b), although 
a spectroscopic identification of the claimed galaxy, along with the 
measurement of its redshift, could not be done because its faintness
and the brightness of the QSO image components preclude this. 
The observed absorption systems along the line of sight at 
$z_{abs} \approx$ 0.61, 1.35, 1.44, 
1.66, 1.87, 2.07 and 2.1 (Monier, Turnshek, \& Lupie 1998) have raised 
the suspicion that the lens might be associated with one or more of the 
known absorption systems.

The first study of gravitational lens models for the Cloverleaf
was based on ground-based optical and VLA observations by Kayser et
al.\ (1990), who found that a singular isothermal elliptical galaxy
model and a two-isothermal-spherical-galaxies model could fit their
optical image positions and some features of the radio data. However,
neither of the Kayser et al.\ (1990) models was consistent with
the observed optical image intensity ratios.  In addition, the
two-spherical-galaxies model predicted an unobserved ``bright''
fifth image.  Kayser et al.\ (1990) considered the possibility
of microlensing to explain the discrepancy between the model
magnification ratios and the observed intensity ratios; either
components A \& D or B \& C should be microlensed according to
their models.  There have been some observational indications for
microlensing of component D in the Cloverleaf (see below and \S 3).  
Photometric variations of component D relative to the other
components were reported in several observations (Kayser et al.\
1990; Remy et al.\ 1996; $\O$stensen et al.\ 1997), although T97
did not detect any significant variations between five different
HST observing epochs covering 2.76 years.  Except for component D, 
however, there are no observational indications for microlensing in 
the Cloverleaf (see also $\O$stensen et al.\ 1997).

More recent modeling attempts for the Cloverleaf were based on HST
archival data (optical image positions and intensity ratios) along
with CO(7-6) emission data (Yun et al.\ 1997; Kneib et al.\ 1998a).
Those authors used lens models to study the CO emitting region
around the central engine of the QSO.  One aspect of particular
interest which is common to the lens models of Yun et al.\ (1997)
and Kneib et al.\ (1998a) is that neither model is a single potential
model. For example, the model of Kneib et al.\ (1998a) considers a
possible contribution from a yet-to-be-confirmed distant cluster along
with a primary lensing galaxy.  These models suggest that the
lens in the Cloverleaf is a complicated lens rather than an isolated
single galaxy.  Keeton, Kochanek, \& Seljak (1997; hereafter KKS) 
argue that many lenses, including the Cloverleaf, can be well-fitted using
{\it two independent shears}, which is also consistent with this
point of view.

In this paper we use the HST WFPC/WFPC2 astrometric and photometric data 
given by T97 to constrain Cloverleaf lens models. In particular, we use the
dust-extinction-corrected relative amplifications as lensing constraints.
We will also use the position of the putative lensing galaxy given by 
Kneib et al.\ (1998b) as a constraint.
We will use the  generalized power-law mass model described 
in Chae, Khersonsky, \& Turnshek (1998; hereafter CKT) to model the lens.  
In \S2 we briefly review the elliptical mass model of CKT
and define the goodness of fit, $\chi^2$, of a model to the data.
In \S3 we investigate gravitational lens models for the Cloverleaf.
We first find that a single galaxy model is inconsistent with the
T97 data.  We then consider lens models consisting of an elliptical
galaxy and an external shear. We also consider lens models consisting of 
two elliptical masses (a primary galaxy + a secondary galaxy; 
a galaxy + a cluster) in order to fit successfully the 
dust-extinction-corrected relative amplifications. 
In \S4 we discuss future observations which can test and/or 
further constrain lens models of the Cloverleaf.

\section{Power-Law Elliptical Mass Model}
The projected view of a power-law triaxial ellipsoid 
on the sky (i.e.\ the lens/image plane) can be represented by the
following 2-dimensional elliptical distribution of mass 
(CKT; see also Keeton \& Kochanek 1997)
\begin{equation}
\Sigma(r,\theta) =
 \Sigma_0 \left\{1+\left(\frac{r}{r_c}\right)^2
           [1+e \cos 2(\theta-\theta_0)] \right\}^{-(\nu-1)/2},
\end{equation}
where $(r,\theta)$ are the polar coordinates of the lens plane. 
The power-law elliptical mass (EM; eq.\ [1]) is described by the following 
five independent parameters: (1) $\Sigma_0$, the surface density at the 
center, (2) $\nu$, the radial index such that the deprojected 3-dimensional
mass density $\rho \sim r^{-\nu}$ and the isothermal-like distribution
is for $\nu = 2$, (3) $r_c$, the core radius, (4) $e$, an ellipticity
parameter which is related to the true ellipticity, $\epsilon$ 
($= 1 - q$ where q is the axis ratio) via
$\epsilon = 1 - [(1-|e|)/(1+|e|)]^{1/2}$, and (5) $\theta_0$, an
orientation angle which is the position angle (P.A., north through east) 
if $e > 0$.

To fit the observed positions and relative amplifications of the image
components (T97) and the observed positions of the lensing galaxy 
(Kneib et al.\ 1998b) we define the goodness of fit, $\chi^2$, as follows
\begin{equation} 
\chi^{2} =
\sum_{i}\left(\frac{x_{iA}^{th}-x_{iA}^{ob}}{\sigma_{x_{iA}}}\right)^{2}
+\sum_{i}\left(\frac{y_{iA}^{th}-y_{iA}^{ob}}{\sigma_{y_{iA}}}\right)^{2}
+\sum_{j}\left(\frac{R_{jC}^{th}-R_{jC}^{ob}}{\sigma_{R_{jC}}}\right)^{2},
\end{equation}
where i = B, C, D, G1 (G1 denoting the primary galaxy) and j = A, B, D.  
The parameters $x_{iA}^{th}$
and $x_{iA}^{ob}$ are the theoretical and observational relative
horizontal positions (i.e.\ right ascensions), respectively, of
components B, C, D, G1 with respect to component A.  The parameters
$y_{iA}^{th}$ and $y_{iA}^{ob}$ are the theoretical and observational
relative vertical positions (i.e.\ declinations), respectively, of
components B, C, D, G1 with respect to component A.  The parameters
$R_{jC}^{th}$ and $R_{jC}^{ob}$ are the theoretical and
observational relative magnifications, respectively, of components
A, B, D with respect to component C. Finally, $\sigma_{x_{iA}}$
and $\sigma_{y_{iA}}$ are the observational uncertainties in the
relative positions, while $\sigma_{R_{jC}}$ are the observational
uncertainties of the relative magnifications.  The parameter $\chi^2$
(eq.\ [2]) is a function of the lens parameters.  Given the positions 
and relative amplifications of the image components and the positions
of the primary galaxy along with their observational uncertainties, 
the parameters of a lens model are adjusted to minimize $\chi^2$. 
For the minimization of $\chi^2$, we use the so-called downhill
simplex method (see Press et al.\ 1992). 

\section{Gravitational Lens Models of the Cloverleaf}

The HST optical images of the Cloverleaf (T97) provide more accurate
and tighter constraints on lens models of the Cloverleaf than any
earlier observations. Except for the declination of component D,
the relative image positions are measured with an accuracy $\approx
3-4$ mas.  Table 1 summarizes the T97 astrometric data. These image
positions will be used as the primary constraints on the lens
models of the Cloverleaf in this study. The photometric data of T97
from the ultraviolet to the near infrared revealed sight-line-dependent 
dust extinctions for the individual image components.  
T97 considered two dust models, galactic-like
and SMC-like, at two assumed redshifts, $z = 2.55$ and $1.55$.
The derived relative amplifications are summarized in Table 2.
Among the four dust-extinction models, Model C (galactic extinction
at $z = 1.55$) is significantly different from the other models which 
are similar to one another. The NICMOS-2/F160W intensity ratios 
(Kneib et al.\ 1998b) now appear to be inconsistent with Model C 
(so this model will no longer be considered).
The main difference between the derived relative amplifications
and the observed intensity ratios in a given filter is that components 
A and B have significantly larger amplifications after corrections
for the dust extinctions.  We will use the dust-extinction-corrected
relative amplifications as lens model constraints in this study.
In previous studies HST optical intensity ratios were used  
(Yun et al.\ 1997; Keeton et al.\ 1997; Kneib et al.\ 1998a).
We will also use the position of the lensing galaxy reported in
Kneib et al.\ (1998b) as a constraint.\footnote{The detected light
could be a fifth image rather than a galaxy. This possibility cannot be
excluded, especially since a spectroscopic identification of the light
has not yet been done. In this paper, however, we will not explore the 
possibility of a fifth image.} This new constraint is
important only for a two-galaxy model (\S 3.3) because a single galaxy
model with and without a perturbation field requires the galaxy to
be positioned near the geometrical center of the Cloverleaf.

However, one caution should be taken in using 
the relative amplifications of T97 as lensing constraints, 
namely any possible contribution from microlensing. 
As well as the observations of photometric variation of component D
mentioned in \S 1, spectroscopic properties in the Cloverleaf 
could be interpreted as an evidence of microlensing of component D. 
In fact, Angonin et al.\ (1990) favored the interpretation that 
two spectral features of component D, 
which were substantially different from the same
features in the other components, were the result of microlensing:
(1) component D has significantly lower broad emission line (BEL) to 
continuum flux ratio than the other components and (2) at some outflow
velocities the BAL profiles of component D are deeper than the
BAL profiles in the other components.  The lower BEL/continuum
ratio of component D than the other components (A : B : C : D = 
0.97 $\pm$ 0.03 : 0.95 $\pm$ 0.03 : 1.00 : 0.70 $\pm$ 0.02)
could be easily explained by microlensing. Namely, the more 
centrally-concentrated continuum-emitting region is amplified (microlensed)
more than the BEL region as seen in component D, thereby lowering its 
BEL/continuum ratio. In line with this interpretation,
several microlensing calculations have been done with assumed
models of the BAL clouds to reproduce the distinctive BAL profiles
of component D (Hutsem\'{e}kers 1993; Hutsem\'{e}kers, Surdej, \&
Van Drom 1994; Lewis \& Belle 1997). 
Given this microlensing interpretation of component D, 
the macrolensing relative amplification of component D can be derived 
by multiplying its observational relative amplification (after
correction for dust extinction) by 0.7. Unless specified otherwise, 
a $\chi^2$ of a model in this paper is given for the amplification 
Model B with the microlensing interpretation of component D, i.e.,
the observational relative amplifications used are
$R_{AC} = 1.68(0.168)$, $R_{BC} = 1.61(0.161)$, 
and $R_{DC} = 0 .665(0.095)$, where the values in parentheses are 
the uncertainties we chose.

Another possibility we
can envisage for the explanation of the lower BEL/continuum ratio
in component D is that the macrolensing magnification ratio of
the BEL/continuum is  smaller for component D than for the other
components.  In this case, there is no need for microlensing in component
D, and the dust-extinction-corrected relative amplifications will
represent the macrolensing magnification ratios. This interpretation
could be interesting because the characteristic radius of the BEL region
could then be estimated from a study of the functional relationship
between the magnification  and radius of the extended region.
This would, of course, require that the lens model for the Cloverleaf
be well constrained.  Another observational constraint from the
T97 observations is that any unresolved image near the geometrical
center of the Cloverleaf should be fainter than $\approx$ 2.5\% of
the brightness of component C.\footnote{If any dust is present along 
the line of sight of a fifth image and thus obscures it,
this limit should be corrected. In this study, we will ignore such a
correction.}  Finally, we will exclude models which
predict very large total magnifications (i.e.\ larger than $\approx$
600) by requiring the QSO to be intrinsically brighter than M$_V$
$\approx -23$, consistent with the spectrum exhibiting BALs.

We calculate the lensing in a standard cosmology with the following
values for the cosmological parameters: the deceleration parameter
$q_0 = \frac{1}{2}$, the Hubble constant H$_0 = 75 h_{75}^{-1}$
km s$^{-1}$ Mpc$^{-1}$, and the cosmological constant $\Lambda = 0$.

\subsection{\it Single Galaxy Lens Models: Isolated}
Kayser et al.\ (1990) considered a singular isothermal elliptical galaxy 
model for the Cloverleaf. This model corresponds to the special case of
$r_c \rightarrow 0$ and $\nu = 2$ in the power-law EM
model described in \S 2. The singular isothermal model predicted
magnification ratios of A : B : C : D = 0.58 : 1.00 : 1.00 :
0.35 when constrained with optical image positions available at
that time.  These ratios are obviously inconsistent with T97's relative
amplifications. Our mass model allows us to consider other values of
$\nu$ and $r_c$. The effect of varying  $r_c$ is related
to the brightness of a fifth image. For a given value of $\nu$,
the upper limit on $r_c$ is set by requiring that any fifth image be
fainter than 2.5\% of the brightness of component C. The low value
of 2.5\% for the brightness limit of any fifth image means that the
upper limit on the core radius is small. For example, the upper limit
on the core radius of a model with $\nu = 1.19$ is only $r_c \approx
0.12 h_{75}^{-1}$ kpc.  Since the image is formed far outside
the core radius (at $r \approx 3.5 h_{75}^{-1}$ kpc), varying core
radius has little effect on the lensing properties of the model.
On the other hand, the model is very sensitive to the value of the
radial index $\nu$. Shallower mass profiles (i.e.\ smaller $\nu$)
require less elliptical (i.e.\ more circular) distributions. Mass
distributions of different ellipticities predict different relative
magnifications, time delays, and total magnifications.  Since the
isothermal model is inconsistent with the observationally-derived
relative amplifications, one is led to vary the radial index from
the isothermal value $\nu=2$.  Using the observational constraints of T97, 
the goodness of fit $\chi^2$ decreases from $\approx 216$
to the minimum value of $\chi_{min}^2 \approx 168$ as the value of $\nu$
decreases from the isothermal value of 2. The minimum $\chi^2$
is obtained when $\nu = 1.19$. For this model the predicted
relative magnifications are A : B : C : D = 1.15 : 1.16 : 1.00 : 0.85 
(see Table 3). 
This is much more consistent with the observed relative amplifications
when compared to the predictions of the isothermal model. However,
it still significantly deviates from the observationally-derived
relative amplifications by T97.  Moreover, although the main difficulty
with the single-galaxy model is the mismatch between the predicted
and observed relative amplifications, it cannot 
reproduce the HST image positions at an acceptable level.
Also, even with less sensitive observational constraints than the
T97 constraints, KKS obtained $\chi^2 = 148.9$ and
142.0 with a single ellipticity and a single external shear, respectively.
Therefore, we conclude that a power-law mass distribution with
elliptical symmetry is inconsistent with the observed properties of
the Cloverleaf.

\subsection{\it Single Galaxy Lens Models:  With External Tidal
Perturbation} In \S 3.1 we saw that a single EM
lens is inconsistent with the observational constraints.
Here we consider the possibility that an external tidal
perturbation is present, in addition to a single primary lensing
galaxy, and that it significantly contributes to the lensing
effect.\footnote{Calculations similar to that presented here can
also be found in KKS, who calculated
lens models for several lenses, including the Cloverleaf.  However,
our calculation for the Cloverleaf is based on the new lensing
constraints of T97.} The tidal perturbation could be due to a galaxy
cluster, of which the lensing galaxy is a member, or a group of
galaxies along the line of sight. We use an external shear term to
describe the tidal perturbation.  The scaled deflection angle of
the external shear term is given by Kovner (1987) as
\begin{equation} \vec{\alpha}_t(\vec{x})
= \gamma \left( \begin{array}{cc} \cos 2\theta_{\gamma} & \sin
2\theta_{\gamma} \\ \sin 2\theta_{\gamma} & -\cos 2\theta_{\gamma}
\end{array} \right) \cdot \vec{x}, 
\end{equation} 
where $\gamma$  and  $\theta_{\gamma}$ 
are the shear strength and angle, respectively, and $\vec{x}$ is
the scaled position vector on the lens plane (see, e.g., CKT).
In order not to increase the number of degrees of freedom of the model
compared with the single EM model, we fix the radial index at 
$\nu = 2$ and the core radius at $r_c = 0.001h_{75}^{-1}$ kpc (even if 
we allowed the parameters $\nu$ and $r_c$ to vary, we would not improve
the fit).

This model of an elliptical mass along with an external shear (to be
denoted ``EM$+\gamma$'' model) can be made to fit the image positions 
within measurement uncertainties.  
However, the model cannot be used to obtain a good fit 
to the relative amplifications. The model has $\chi^2 \approx 12.8$.
The fitted model parameters and model predictions 
can be found in Table 3. Note that the model 
predictions of $R_{BC}$ and $R_{DC}$ do not match well 
the observationally-derived relative amplifications. The galaxy
has a P.A. of $\approx 140\deg$, so it is oriented roughly along BC.
The shear angle is $\theta_{\gamma} \approx 44\deg$, 
which puts its orientation roughly along AD (orthogonal to the galaxy).
If the galaxy is forced to be oriented along AD, then the fitted
shear angle tends to be aligned with the galaxy, and the fit becomes much
worse. Such a model becomes qualitatively similar to the single EM model 
considered in \S3.1. Thus, in using EM$+\gamma$ model to fit the 
Cloverleaf, it is critical for the  galaxy to be oriented along BC.
In previous studies, Yun et al.\ (1997) and KKS also 
found that an elliptical potential and an elliptical mass, respectively, 
would be oriented roughly along BC, while a shear would be oriented
roughly along AD, consistent with our result. The model predictions
of Yun et al.\ (1997) do, however, differ somewhat from our model
predictions; their model predicts ${\cal M}_{tot} = 220$ and maximum
time delay $\tau_{DC} = 1.9 h_{75}^{-1}$ days. The fitted parameters of
the KKS model and our model are similar, and the quantitative difference
between the two results comes from the fact that we use the 
dust-extinction-corrected relative amplifications by T97 as constraints. 
We note that a shear of $\gamma \approx 0.22$ required in the EM$+\gamma$ 
model is somewhat large, and such a strong shear implies an important
secondary lensing contributor (see KKS). 
 
\subsection{\it Two-Galaxy Lens Models}
In \S3.2, we saw that an EM$+\gamma$ model does not fit well the 
observationally-derived relative amplifications (in particular $R_{BC}$). 
Here we consider the possibility of the presence of a secondary galaxy 
near the line of sight, either at the same redshift or at 
a different redshift from the primary galaxy.
The redshift(s) of the lensing object(s) is(are) unknown in the Cloverleaf. 
We consider the case that two galaxies are at the same redshift and 
arbitrarily set $z = 1.55$. A lens model consisting of two elliptical
masses (to be denoted ``EM2'' model) has twice as many free parameters as 
a single-galaxy lens model considered in \S 3.1. 
When we describe each galaxy by an EM model, 
it is characterized by five parameters (see \S 2). In
addition to the ten parameters of two galaxies, there are four more
free parameters in the EM2 model, i.e., the relative position 
of a galaxy with respect to the other on the lens plane and the two
coordinates of the QSO on the source plane.

From now on we will assume that the two galaxies have the same
radial index and core radius. Furthermore, we fix the core radius at
$r_c = 0.001 h_{75}^{-1}$ kpc and the radial index at $\nu = 2$, which
is the isothermal distribution. In the two-galaxy model a fifth image 
can be formed between the two galaxies, and the relative brightness of 
the fifth image depends on the core radius of the galaxy closest to the 
fifth image and the mass ratio between the two galaxies. 
Given the constraint that any fifth image should be
fainter than 2.5\% of the brightness of component C (T97), it would
be reasonable to fix the core radius at a small value. There is no
{\it a priori} reason that the two galaxies should have the same radial
index.  We have chosen to simplify our analysis by assuming so. 
 With the above assumptions (simplifications)
the parameters of the two-galaxy model are 
(1) the combined central density, $\Sigma_0^{tot}$ 
($=\Sigma_0^{(1)} + \Sigma_0^{(2)}$), (2) the mass fraction of G2, $f_2$ 
($=\Sigma_0^{(2)}/\Sigma_0^{tot}$), (3) and (4) 
the ellipticities of the two galaxies, $\epsilon_1$ and $\epsilon_2$, 
(5) and (6) the position angles of the two galaxies, P.A.$_1$ and P.A.$_2$,
(7) and (8) the magnitude ($d_{12}$) and position angle (P.A.$_{12}$) of 
the displacement of G2 from G1 ($\vec{d}_{12}$), and (9) and (10) the
two coordinates of the QSO on the source plane. In the above, G1 is
the primary galaxy closest to the geometrical center of the Cloverleaf,
while G2 is the secondary galaxy. 

\subsubsection{\it Secondary Galaxy at an Arbitrary Position}
We first consider the possibility of the presence of an undetected
secondary galaxy in the vicinity of the Cloverleaf. We find that 
the two-galaxy model with G2 at an arbitrary position
can fit both the QSO image positions and relative amplifications 
nearly perfectly. However, the position of the primary galaxy is 
not fitted as well. The value of $\chi^2$ (ranging $\approx 5.4 - 7.7$) 
is dominated by the mismatch between the model (primary) galaxy position 
and the observational galaxy position obtained by Kneib et al.\ (1998b). 
In fact, if we were not using the galaxy position as a constraint, 
we would have a perfect fit. The secondary galaxy can be positioned 
(1) southeast of the primary galaxy within the Einstein ring 
[Model 1, Table 4; Fig.\ 1 (a)], (2) north of component 
D [Model 2, Table 4; Fig.\ 1 (b)], or (3) southwest of the primary galaxy 
outside the Einstein ring [Model 3, Table 4; Fig.\ 1 (c)]. 
In all of the above cases the secondary galaxy
is significantly less massive than the primary galaxy ($f_2$ ranges from
$0.17 - 0.26$) so that the secondary galaxy could be even fainter than
the primary galaxy which has not been detected at optical wavelengths. 
The projected separation ($d_{12}$) between 
the two galaxies is not uniquely determined, and varying its value
does not lead either to a significant change in $\chi^2$ or to different  
lensing predictions (e.g.\ time delays, total magnification). 
Thus, for each model we fixed the parameter $d_{12}$ at an (arbitrarily) 
chosen value. As Figure 1 (a) shows, in Model 1 an extended region can 
form a weak fifth image near the center of G2. However, its relative  
magnification (with respect to component C) is only 
$\approx 9 \times 10^{-4}$. In the above two-galaxy lens models 
the primary galaxy is oriented roughly along AD. As we will see, 
the two-galaxy model with an arbitrary position for G2 matches 
the T97 observational constraints better than any of the other models
considered in this paper. This is, of course, the result of taking advantage
of the freedom in G2's position. As seen above, G2's position cannot be
uniquely determined using lensing constraints. However, there are 
only three regions of sky [see Figure 1 (a), (b), and (c)] in the vicinity 
of the Cloverleaf which should be considered
candidates for where a faint galaxy could be found. Next we consider
the two-galaxy model with G2's position constrained.

\subsubsection{\it Secondary Galaxy at the Position of a Detected Object}
We examine the possibility that the secondary galaxy is
one of the detected objects in the surrounding of the Cloverleaf
(see Kneib et al.\ 1998a,b). Specifically, we consider their object number 
14, which is $\approx 4''.5$ north-northeast of component A. 
From the NICMOS-2 data we derive ($1''.87$, $4''.14$) as the position 
of the object relative to component A with an error of  $0''.07$. 
However, the object is not bright enough to derive its light distribution.
Even if it were possible to derive the light distribution, we could not use 
it as the mass distribution since the distribution of dark mass,
which is likely to account for most of the total mass in the scale of
$4''.5$ ($\approx 26h_{75}^{-1}$ kpc), could be significantly different from
the light distribution. Thus, we allow the ellipticity and position angle of
G2 to be free parameters. This model can fit well both the image positions
and relative amplifications, except for $R_{DC}$. The predicted $R_{DC}$ 
($0.31-0.56$) is significantly lower than the observational $R_{DC} = 0.67$, 
which was derived using the microlensing interpretation of the 
{\it continuum} of component D from the dust-extinction-corrected relative
amplification. However, if the BELR is also microlensed in component D, 
the macrolensing $R_{DC}$ should be lower than 0.67 and its value would 
depend on the microlensing amplification factor. Since the microlensing 
amplification depends on the size scale of the source region, it will be 
different for the continuum region and the BELR; the smaller 
continuum region will be more amplified than the more extended BELR.
In this case, the discrepancy of $R_{DC}$ can be clearly resolved. For 
this possibility, in this section (and only in this section) we do not use 
$R_{DC}$ as a constraint. Instead, we will derive the required microlensing
amplification factors for the continuum {\it and} the BELR using the model.

In the model G2 has to be highly elliptical and the best fitting model is 
obtained when G2's ellipticity is $\epsilon_2 = 0.84$ 
[Model 4, Table 4; Fig.\ 1 (d)]. As $\epsilon_2$ decreases from this value, 
$\chi^2$ increases rapidly. For a 99\% confidence limit 
$\epsilon_2 \gtrsim 0.58$. As $\epsilon_2$ increases from the best fitting
value, $\chi^2$ increases only moderately. The largest ellipticity we
consider is $\epsilon_2 = 0.94$. The predicted $R_{DC}$ and required 
microlensing amplification factors of the continuum and BELR are shown 
in Figure 2 as a function of $\epsilon_2$ within the 99\% confidence limit.
In the model the primary galaxy has a relatively small ellipticity and 
is oriented roughly along BC, as opposed to the other two-galaxy models
considered in \S 3.3.1. 

Although the model considered in this section is a good candidate for the 
correct lens model of the Cloverleaf, it is still possible that the BELR is
extended enough so that it is unaffected microlensing. In this case
the discrepancy of $R_{DC}$ should be explained in other ways (e.g.\ G2 may
be at a different redshift from G1), or the model should be discarded. 
The model could also be discarded if it could be determined that G2 is less 
massive and/or less elliptical than required in the model 
[in the model the total mass of G2 within $4''.5$ is $\approx (0.66-3.3) 
\times 10^{12}$ M$_{\odot}$ for $0.58 \lesssim \epsilon_2 \lesssim 0.94$]. 
At present we do not know if the model is the true description of the lens. 
This is a posteriori reason that we explore other models 
(\S 3.3.1 and \S 3.4).

\subsection{\it Galaxy+Cluster Lens Models: A Model-Dependent Bound on 
the Mass of the Lensing Galaxy} 
In this section we re-examine the possibility
that a cluster is present and makes a significant contribution to the
lensing. Although an external shear term considered in \S 3.2 could
qualitatively account for the lensing contribution of the cluster, the 
ignored higher-order terms (rather than the quadrupole term) may be 
important. In this case one must model the cluster, for example, 
using an elliptical mass distribution. Since the properties of the possible 
cluster are unknown at present, we make several assumptions about the 
mass distribution of the cluster for studying its lensing effect. 
We consider an isothermal mass distribution ($\nu = 2$) for the cluster
and fix the core radius at $r_c = 40h_{75}^{-1}$ kpc.
The chosen core radius is similar to the value of 33$h_{75}^{-1}$ kpc 
used by Kneib et al.\ (1998a). It is also in the same order of magnitude
as the isothermal core of 23$h_{75}^{-1}$ kpc for the cluster 0957+561
as determined from weak lensing effect by Fischer et al.\ (1997). 
We fix the ellipticity of the cluster at $\epsilon = 0.5$. 
Buote \& Canizares (1996) derive $\epsilon_{mass} \approx 0.40-0.55$ 
for isothermal mass models using ROSAT observations of five Abell clusters 
and assuming hydrostatic equilibrium. 
De Theije, Katgert, \& van Kampen (1995) find that
the ellipticity peaks at $\epsilon \sim 0.4$ and extends to 
$\epsilon \sim 0.8$ from an analysis of a large set of 99 low-redshift
Abell clusters. In the above, however, the exact values of $r_c$ and
$\epsilon$ are less important for our purpose since the lensing effect
is not very sensitive to these parameters.
Kneib et al.\ (1998a) found that an overdensity of objects around the 
Cloverleaf could be centered $\approx 7''.8$ southwest of component A. 
Consistent with their finding, 
we fix the separation between the galaxy mass center and the cluster mass 
center at $d_{12} = 8''$. However, we allow the position angle of the cluster
from the galaxy, P.A.$_{12}$, to be a free parameter. The exact value of
$d_{12}$ is less important since varying $d_{12}$ does not lead 
either to a significantly-improved fit or to different lensing predictions.
On the other hand, varying the parameter P.A.$_{12}$ can lead to 
a significant change in $\chi^2$ and also to different lensing predictions. 
The parameter P.A.$_{12}$ can be varied by varying the parameter $f_2$,
since the fitted value of P.A.$_{12}$ is determined by the chosen
value of $f_2$. The parameter $f_2$ determines the masses of the lensing
galaxy and the cluster. If the cluster is relatively more massive, then
the galaxy is relatively less massive. In this case $f_2$ is relatively
larger. This is important because the lensing galaxy can be
significantly less massive compared with the case where there is no
cluster contribution. A best fitting model has $\chi^2 \approx 6.25$ 
and is obtained when $f_2 \approx 0.0011$, 
for which P.A.$_{12} \approx -98\deg$ (Model 1, Table 5; Fig.\ 3). 
As $f_2$ increases from this value, $\chi^2$
increases only moderately. However, for $f_2 \gtrsim 0.0025$,
the total magnification is ${\cal M}_{tot} \gtrsim 600$.
Thus, $f_2 = 0.0024$ is the upper limit, since we require the total 
magnification not be larger that 600 (see the introduction of \S 3).
The model with $f_2 = 0.0024$ is given in Table 5 (Model 2).
As $f_2$ decreases from 0.0011, $\chi^2$ increases rapidly.
We set the lower limit on $f_2$ at the 99\% confidence level from
the best fitting model. The lower limit is $f_2 = 0.0003$ 
(Model 3, Table 5). For the range of $f_2 = 0.0003$ to 0.0024 
the mass of the lensing galaxy inside $0''.6$ (which is approximately
the Einstein radius) ranges from $0.78\times 10^{11} {\mbox M}_{\odot}$ 
to $1.84\times 10^{11} {\mbox M}_{\odot}$ with the best fitting value of
$1.22\times 10^{11} {\mbox M}_{\odot}$.  Our best fitting model 
(Model 1) is much more consistent with the observational
relative amplifications than is the Kneib et al.\ (1998a) galaxy+cluster model
(their Model 2). This improvement was achieved only when we allowed the 
parameter P.A.$_{12}$ to be a free parameter. The best fitting value of 
$-98\deg$ is significantly different from the value of $-144\deg$ for 
the Kneib et al.\ (1998a) model (the cluster center is more westward in our 
model). This is, of course, not an issue since no reliable determination of 
the mass center of the possible cluster is known at present. Rather, our 
study suggests that a cluster (if it is truly present) should be positioned 
$\sim$ 46$\deg$ more westward than Kneib et al.\ (1998a) found.

\subsection{\it BEL/Continuum Ratio and a Size-Scale Constraint
on the BEL Region}  Finally, we turn to the question of the lower
BEL/continuum ratio for component D.  Using our lens models 
(EM$+\gamma$, two-galaxy, and galaxy+cluster models),
we have examined the macrolensing interpretation of the
lower BEL/continuum ratio for component D. To consider this we
studied the magnification of a circular region on the source plane.
A lensed image of a circular source will not be circular, but
will have a deformed shape, and its angular area will be different
from that of the source.  We studied how the angular areas of
the image components vary as the radius of the source increases.
We found that for our lens models, the ratios of the angular areas
of the image components changed little up to the point where two
of the components merged (components A and B merged first). In particular,
the ratio of the angular area of component D to that of component C
does not decrease as the radius increases. Even if an elliptical source 
is considered, it does not make any significant difference.
Thus, the lower BEL/continuum ratio for component D cannot be explained
by macrolensing using our lens models.

As was noted in the beginning of \S3, the most plausible interpretation 
is that component D is microlensed. Since microlensing amplification 
depends on the size of the source region, the pointlike continuum region 
and the more extended BEL region will be amplified differently. 
While we can be confident that the continuum region is amplified by a small 
mass, we do not know to what degree the BELR is amplified by the small mass
since we do not know the extent of the BELR. Whether or not 
a source region will be affected by a microlens is determined by the 
characteristic length $\eta_0$ on the source plane 
(Schneider, Ehlers, \& Falco 1992). For a star of mass $M$ at $z=1.55$, 
it is given by 
\begin{eqnarray}
   \eta_0 & \approx & 
0.0061 (M/M_{\odot})^{1/2} h_{75}^{-1/2} \mbox{pc} \nonumber \\
      &  \approx &
 1.9 \times 10^{16} (M/M_{\odot})^{1/2} h_{75}^{-1/2} \mbox{cm} , 
\end{eqnarray}
where we take $q_0 = \frac{1}{2}$ and $\Lambda = 0$.
If the BELR is larger than $\eta_0$, it will not be significantly
affected by microlensing. Otherwise, the BELR will be amplified, as will
the continuum. In using $R_{DC}$ as a lensing constraint in \S3.1 $-$
\S3.4, we assumed that the BELR was not amplified by a microlens.
However, this assumption was excluded when we considered the 
two-galaxy lens model with the secondary galaxy at the position of 
object 14 in \S3.3.2. It is therefore interesting that the lens model
of \S3.3.2 requires that the BELR be amplified. This is important
because if the model is correct, and all the other models could be rejected
observationally (e.g.\ the proposed secondary galaxies in \S3.3.1 turn out 
not to exist and the possible cluster
in \S3.4 does not have a significant lensing effect because it is not
sufficiently massive), the BELR must be as large 
as, or {\it smaller} than, $\eta_0$. A detailed study of microlens models
could then be used to determine or constrain the physical size of
the BELR.

The likelihood of microlensing along a light path is determined by 
the ``optical depth'' to microlensing, which is given by the ratio of 
surface mass density of microlensing matter to the critical mass density 
$\Sigma_{cr} = (4 \pi G /c^{2})^{-1} ( D_{d} D_{ds}/D_s)^{-1}$ 
(Paczy\'{n}ski 1986). The two-galaxy models have qualitatively different 
microlensing probabilities for the image components compared with the 
EM$+\gamma$ model and the galaxy+cluster models. In the two-galaxy lens models 
(from both \S3.3.1 and \S3.3.2) component D has the largest microlensing 
probability, while in the other models component D has a slightly smaller 
probability than component B and/or C. More specifically, component D has 
$\approx 6-31$\% higher microlensing probability than component A (or C) 
which has the second largest microlensing probability in the two-galaxy
lens models. Thus, the microlensing interpretation of component D 
favors the two-galaxy model over the single-galaxy, external-shear 
model and the galaxy+cluster model.

\section{Discussion} 
We have argued that a single-galaxy model is inconsistent with the HST 
observations (T97). While the introduction of an external shear to 
a single galaxy can fit the image positions well, the predicted relative
magnifications do not match well the observationally-derived relative 
amplifications (\S 3.2). Two possibilities have been considered to explain 
this. One possibility is that none of the dust extinction models used by 
T97 to derive the relative amplifications accurately model the true dust 
extinctions, i.e., the dust extinctions might be more complicated than 
assumed by T97. Note that T97 had to assume the type and redshift of the 
dust since they were not known from observations. Because of this, we were
conservative in using the relative amplifications of T97 as lensing 
constraints; we used 10\% as uncertainties. Yet, the single-galaxy, 
external-shear model is still not very successful in matching the 
amplifications. 
Moreover, since the NICMOS-2/F160W intensity ratios (Kneib et al.\ 1998b) 
are consistent with the T97 amplification models A, B, and D (Table 2), which
are similar to one another, this bolsters our conclusion that a single-galaxy,
external-shear model is not an appropriate description of the lens in
the Cloverleaf, although the model is relatively simple. In other words,
the quadrupole term (eq.\ [3]) alone may not appropriately account for 
a secondary lensing contribution. 

The other possibility we considered were new lens models of the Cloverleaf.
Specifically, we considered lens models consisting of two elliptical masses; 
two-galaxy models (\S 3.3) and galaxy+cluster models (\S 3.4). In the 
two-galaxy models we considered a secondary galaxy either at an arbitrary
position (\S3.3.1) or at the position of a detected object (\S3.3.2).
The two-galaxy models of \S3.3.1 can fit the observationally-derived relative 
amplifications nearly perfectly. The two-galaxy model of \S3.3.2 can fit 
$R_{AC}$ and $R_{BC}$ well, but its predicted $R_{DC}$ is significantly
lower than the observed $R_{DC}$. The galaxy+cluster model does not 
fit the relative amplifications as well as the two-galaxy models of \S3.3.1,
but it does fit somewhat better than the single-galaxy, external-shear model. 
In the two-galaxy models of \S3.3.1 the secondary galaxy is much less massive 
(and thus much fainter, assuming a similar mass-to-light ratio) than the 
primary galaxy. Given that the primary galaxy is so faint that it has never
been detected at optical wavelengths, it may not be a problem that
the secondary galaxy has not been detected. Even with future observations
it will be a challenge to detect a relatively faint galaxy in the bright
region of the Cloverleaf. As our study shows, the position of the secondary 
galaxy is not uniquely determined, but we proposed three possible regions 
of sky for the unseen galaxy [see Figure 1 (a), (b), and (c)].
The assumption behind these two-galaxy models is that neither galaxy 14 
nor a possible cluster makes a significant contribution to the lensing; 
in other words, we need an unseen lensing contributor. Along the 
line-of-sight to the Cloverleaf, absorption systems exist at many different 
redshifts (Monier et al.\ 1998), and two galaxies may happen to lie along 
the line-of-sight. The two galaxies could be at different redshifts
independent of each other, or at nearly the same redshift. 
For example, they could be a binary system of galaxies, or two galaxies 
in the process of merging. There are at least two other lens systems where
it is believed/speculated that the lens may be two galaxies, namely 
MG2016+112 (Lawrence et al.\ 1984; Nair \& Garrett 1997) and 
B1608+656 (Myers et al.\ 1995; see, however, Fassnacht et al.\ 1996). 
One possible difficulty with the two-galaxy model at present is that 
the predicted position of the primary galaxy is not in good agreement 
with the observed position of the detected light (Kneib et al.\ 1998b). 
However, none of the models considered in this paper
match the observed galaxy position within the observational uncertainty.
The discrepancies could be explained by the misalignments between the
light and mass.\footnote{More radically, the light could be a ghost image
arising from the four bright QSO components. This possibility should not
be excluded until the properties of HST NICMOS images are better quantified.
In this case, a primary galaxy in the two-galaxy model can be near components
A \& B while a secondary galaxy is near components C \& D as opposed to
Model 1 [Fig.\ 1 (a)].} 
The galaxy+cluster lens model for the Cloverleaf was first 
considered by Kneib et al.\ (1998a) based on the detection of the 
overdensity of objects in the surrounding of the Cloverleaf. As was shown
in our study, if the cluster position is fixed at the position used by
Kneib et al.\ (1998a), the model does not fit the observed
relative amplifications very well. However, if we allow the position angle of
the cluster mass center (viewed from the lensing galaxy) to be a free
parameter, we can improve the fit significantly. The best fitting model
predicts that the cluster mass center is positioned roughly west of the
Cloverleaf (P.A.$_{12} \approx -98\deg$). (In the Kneib et al.\ 1998a
model, the cluster is positioned south-west of the Cloverleaf.)
The two-galaxy model with the secondary galaxy's position constrained 
(\S3.3.2) is more realistic than other models in that an {\it observed} 
galaxy is used as the secondary galaxy, while in the other models the 
secondary lensing contributors are either unconfirmed or hypothetical.
However, in order for the model to be consistent with the 
dust-extinction-corrected relative amplifications, component D must be 
microlensed. Furthermore, not only the continuum but also the BEL should 
be amplified by microlensing. This requirement of the model is not  
discrepant with present observations since microlensing of component D 
is very likely (\S1, \S3) and the physical size of the BELR is unknown.

As was shown in \S3.5, the microlensing interpretation of component D
favors the two-galaxy models over the galaxy+cluster models since 
component D has the largest microlensing probability in the two-galaxy 
models, while it does not in the galaxy+cluster models. Kneib et al.\
(1998a) argued that  a cluster is present along the line
of sight to the Cloverleaf. However, its mass distribution is not 
known at present, and thus we do not know how much of a lensing effect 
it will have on the Cloverleaf. Future observational studies 
(e.g.\ using AXAF) of any cluster will be useful for studying its 
lensing effect. 

A long-period (1987-1994) monitoring program of the Cloverleaf
by $\O$stensen et al.\ (1997) did not result in the detection
of time delays between image components. However, upper limits 
of 150 days (99\% confidence) were put on the time delays.  This is
consistent with our lens models.  Measurements of time delays from
future monitoring of the Cloverleaf will be useful to distinguish
between models. Two-galaxy models predict that component C is 
the leading image, followed by components B, A, and D, which is the 
trailing image. The maximum time delay between the leading and trailing 
images ranges between $\approx 20 - 40$ days. In the galaxy+cluster lens 
models component A may lead component B; however, components C and D are
still the leading and trailing images. The maximum time delay is
$\approx 7 - 13$ days.

More work will have to be done to determine if the Cloverleaf
is a good candidate for determining the Hubble constant.
To determine the Hubble constant, time delays between the image
components must be measured and the mass distribution of the
lens must be well-determined, consistent with observational
constraints.  At present, neither of these conditions has been
met for the Cloverleaf.  As our study shows, time delays in the
Cloverleaf are likely to be neither too short nor too large to raise
observational difficulties.\footnote{See, however, Yun et
al.\ (1997) who predicted very short time delays of $\sim$ 1 day.}
For the determination of the mass distribution of the lens, it should be 
a priority to  spectroscopically identify  the reported lensing galaxy 
and observe its properties in detail. It will also be useful to 
observationally study ``galaxy 14'' and the possible cluster. 
Another important observational task is to revisit the relative 
amplification models of T97 in order to determine the macrolensing 
magnification ratios more reliably and accurately. This could be done
with better infrared data.
With observational knowledge of the
 lens and better data on the lensing properties in the future, 
the mass distribution of the lens could be well-determined.

\bigskip

We thank Valery Khersonsky for discussions on mass models of lenses in 
the early stages of this work and Cyril Hazard for reading the early
draft of the manuscript and commenting on it. We also thank Jean-Paul Kneib
for valuable comments and suggestions for the improvement of
the manuscript. K. H. C. acknowledges partial support from
the Zaccheus Daniel predoctoral fellowship at the University
of Pittsburgh during some phase of this work.

\newpage

\centerline{\bf Table 1}
\centerline{\bf HST Astrometric Data}
\tabskip=4.0em
\halign to \hsize{\hfil#\hfil&\hfil#\hfil&\hfil#\hfil
\cr
\noalign{\vskip6pt\hrule\vskip3pt\hrule\vskip6pt}
Component & $\Delta$RA (err) & $\Delta$Dec (err) \cr
          &  (arcsec)        &  (arcsec) \cr
\noalign{\vskip6pt\hrule\vskip6pt}
Comp A &  0.000  &  0.000 \cr
Comp B &  0.744 (0.003) & 0.172 (0.003) \cr
Comp C & -0.491 (0.003) & 0.716 (0.004) \cr
Comp D &  0.355 (0.003) & 1.043 (0.012) \cr
G1$^1$  &  0.12 (0.040) & 0.50 (0.040) \cr
\noalign{\vskip3pt\hrule}
}
$^1$ This position of the primary galaxy is from the analysis of HST 
NICMOS-2 data (Kneib et al.\ 1998b).

\centerline{\bf Table 2}
\centerline{\bf ``Dust-Extinction-Corrected'' Relative Amplifications 
based on HST photometry}
\tabskip=2.0em
\halign to \hsize{\hfil#\hfil&\hfil#\hfil&\hfil#\hfil&\hfil#\hfil&\hfil#\hfil
\cr
\noalign{\vskip6pt\hrule\vskip3pt\hrule\vskip6pt}
Relative & Model A$^1$ & Model B$^2$ & Model C$^3$ & Model D$^4$ \cr
Amplifications$^5$ &    &   &  &  \cr
\noalign{\vskip6pt\hrule\vskip6pt}
$R_{AC}$ & 1.65 & 1.68 & 2.21 & 1.63 \cr
$R_{BC}$ & 1.62 & 1.61 & 2.31 & 1.55 \cr
$R_{DC}^6$ & 0.95 & 0.95 & 1.01 & 0.95 \cr
\noalign{\vskip3pt\hrule}
}
$^1$ Results assuming Galactic Extinction at $z = 2.55$

$^2$ Results assuming SMC Extinction at $z = 2.55$

$^3$ Results assuming Galactic Extinction at $z = 1.55$

$^4$ Results assuming SMC Extinction at $z = 1.55$

$^5$ The adopted uncertainties we use in the lens models are 10\% of these
values.

$^6$ This represents the amplification before the possible microlensing
amplification is corrected (see \S 3). To correct for the microlensing
amplification of the continuum, this ratio should be multiplied by 0.7.

\newpage
\centerline{\bf Table 3}
\centerline{\bf Single-Galaxy Lens Models}
\tabskip=4.0em
\halign to \hsize{#\hfil&\hfil#\hfil&\hfil#\hfil
\cr
\noalign{\vskip6pt\hrule\vskip3pt\hrule\vskip6pt}
          & EM Model & EM$+\gamma$ Model \cr
\noalign{\vskip6pt\hrule\vskip6pt}
$\chi^2$    &  168.3    & 12.76 \cr
N$_{dof}$   &   4     &  4    \cr
$\Sigma_0$ (M$_{\odot}$ pc$^{-2}$)& $2.95\times 10^4$ & $1.13\times 10^7$ \cr 
$\nu$     &  1.19   & 2.00 (fixed)  \cr
$r_c$ ($h_{75}^{-1}$ kpc) &  0.001 & 0.001 (fixed) \cr
$\epsilon$ & 0.059  & 0.425   \cr
P.A. ($\deg$)  & 22.0  & 140.3  \cr
$\gamma$   & --- & 0.217  \cr
$\theta_{\gamma}$ ($\deg$) & --- & 43.6 \cr
\noalign{\vskip6pt\hrule\vskip6pt}
Predictions &           &       \cr
\noalign{\vskip6pt\hrule\vskip6pt}
${\cal M}_{tot}$ &  373  &  36.3  \cr
$R_{AC}$    &  1.15  &  1.56  \cr
$R_{BC}$    &  1.16  &  1.35  \cr
$R_{DC}$    &  0.85  &  0.84  \cr
$\tau_{AC}$ ($h_{75}^{-1}$ days) & 3.93 & 7.38  \cr
$\tau_{BC}$ ($h_{75}^{-1}$ days) & 1.28 & 2.61  \cr
$\tau_{DC}$ ($h_{75}^{-1}$ days) & 6.36 & 15.9  \cr
$\vec{x}_{AG1}$ $('','')$ & (0.165, 0.515) & (0.157, 0.547) \cr
\noalign{\vskip3pt\hrule}
}

\newpage
\centerline{\bf Table 4}
\centerline{\bf Two-Galaxy Lens Models}
\tabskip=1.2em
\halign to \hsize{#\hfil&\hfil#\hfil&\hfil#\hfil&\hfil#\hfil&\hfil#\hfil
\cr
\noalign{\vskip6pt\hrule\vskip3pt\hrule\vskip6pt}
        & Model 1 &  Model 2 & Model 3 & Model 4$^{1,2}$ \cr
\noalign{\vskip6pt\hrule\vskip6pt}
$\chi^2$ &  5.43  & 7.61 & 7.69  & 6.39  \cr
N$_{dof}$  & 2 & 2  & 2  &  2  \cr
$\Sigma_0^{tot}$ (M$_{\odot}$ pc$^{-2}$) 
   & 1.204$\times 10^7$ & 1.298$\times 10^7$ 
   & 1.417$\times 10^7$ &  1.617$\times 10^7$ \cr
$f_2$   & 0.174  & 0.211 & 0.260 & 0.351 \cr
$\epsilon_1$ & 0.340  & 0.340 & 0.355 & 0.172 \cr
P.A.$_1$ ($\deg$)  & 23.6 & 49.6 & 0.5 & 124.2 \cr
$\epsilon_2$ & 0.540 & 0.796 & 0.572  & 0.84 \cr
P.A.$_2$ ($\deg$)  & 36.9  & -6.9 & 43.1 & 31.5 \cr
$d_{12}$ ($''$)  & 0.5 (fixed) & 1.3 (fixed) & 1.6 (fixed) & 3.96 \cr
P.A.$_{12}$ ($\deg$) & 130.0 & -4.5 & -129.5 & 24.7  \cr
$\vec{x}_{G1S}$ $('','')$ & (0.050, -0.076) & (-0.026, 0.090) 
 & (-0.167, -0.128) & (0.022, 0.284) \cr
\noalign{\vskip6pt\hrule\vskip6pt}
Predictions   &   &  &  & \cr
\noalign{\vskip6pt\hrule\vskip6pt}
${\cal M}_{tot}$  &  22.6  & 17.9 & 17.0 & 41.8 \cr
$R_{AC}$   & 1.64  & 1.68  & 1.63 & 1.78  \cr
$R_{BC}$  & 1.60 & 1.59  & 1.62  & 1.58 \cr
$R_{DC}$  & 0.70 & 0.66 & 0.68 & 0.43 \cr
$\tau_{AC}$ ($h_{75}^{-1}$ days)  &  15.9  & 17.8 & 19.7 & 6.9 \cr
$\tau_{BC}$ ($h_{75}^{-1}$ days)  & 8.6 & 9.4  & 8.7  & 3.5 \cr
$\tau_{DC}$ ($h_{75}^{-1}$ days)  & 25.7 & 40.8 & 36.1 & 21.9 \cr
$\vec{x}_{AG1}$ $('','')$ & (0.105, 0.590)  & (0.216, 0.553) 
 & (0.214, 0.557)  & (0.205, 0.544) \cr
\noalign{\vskip3pt\hrule}
}

$^1$ In this model the position of G2 is constrained by 
the observed position of object 14 (see \S 3.3.2).

$^2$ In this model $R_{DC}$ was not used in the calculation of $\chi^2$.

\newpage
\centerline{\bf Table 5}
\centerline{\bf Galaxy+Cluster Lens Models}
\tabskip=3.0em
\halign to \hsize{#\hfil&\hfil#\hfil&\hfil#\hfil&\hfil#\hfil
\cr
\noalign{\vskip6pt\hrule\vskip3pt\hrule\vskip6pt}
        & Model 1 &  Model 2 & Model 3 \cr
\noalign{\vskip6pt\hrule\vskip6pt}
$\chi^2$ &  6.25  & 10.66 & 16.18 \cr
N$_{dof}$   & 3 & 3 & 3 \cr
$\Sigma_0^{tot}$ (M$_{\odot}$ pc$^{-2}$) 
   & 5.52$\times 10^6$ & 3.53$\times 10^6$  & 8.38$\times 10^6$ \cr
$f_2$   & 0.0011  & 0.0024 & 0.0003\cr
$r_c^{(1)}$ ($h_{75}^{-1}$ kpc) & 0.001 (fixed) & 0.001 (fixed) 
& 0.001 (fixed) \cr
$\epsilon_1$ & 0.360  & 0.333 & 0.305 \cr
P.A.$_1$ ($\deg$)  & 135.7 & 132.6 & 139.4 \cr
$r_c^{(2)}$ ($h_{75}^{-1}$ kpc) & 40 (fixed) & 40 (fixed) 
& 40 (fixed) \cr
$\epsilon_2$ & 0.5 (fixed) & 0.5 (fixed) & 0.5 (fixed) \cr
P.A.$_2$ ($\deg$)  & 17.2  & 26.3 & 35.3\cr
$d_{12}$ ($''$)  & 8 (fixed) & 8 (fixed) & 8 (fixed) \cr
P.A.$_{12}$ ($\deg$) & -98.2 & -68.8 & -135.1  \cr
$\vec{x}_{G1S}$ $('','')$ & (-6.621, 0.345) & (-8.892, 3.876) & 
 (-1.421, -1.058) \cr
\noalign{\vskip6pt\hrule\vskip6pt}
Predictions   &   &   & \cr
\noalign{\vskip6pt\hrule\vskip6pt}
${\cal M}_{tot}$ & 185.2 & 583.4 & 60.6 \cr
$R_{AC}$         & 1.52  & 1.44  & 1.28\cr
$R_{BC}$         & 1.67  & 1.67  & 1.32\cr
$R_{DC}$         & 0.76  & 0.58 & 0.84 \cr
$\tau_{AC}$ ($h_{75}^{-1}$ days)  & 2.96  & 4.86 & 7.07 \cr
$\tau_{BC}$ ($h_{75}^{-1}$ days)  & 4.11  & 6.80 & 3.68 \cr
$\tau_{DC}$ ($h_{75}^{-1}$ days)  & 11.2 & 6.74  & 13.0 \cr
$\vec{x}_{AG1}$ $('','')$ & (0.193, 0.535)  & (0.220, 0.519) 
                          & (0.171, 0.548) \cr
\noalign{\vskip3pt\hrule}
}

\newpage

\begin{figure}
\centerline{\psfig{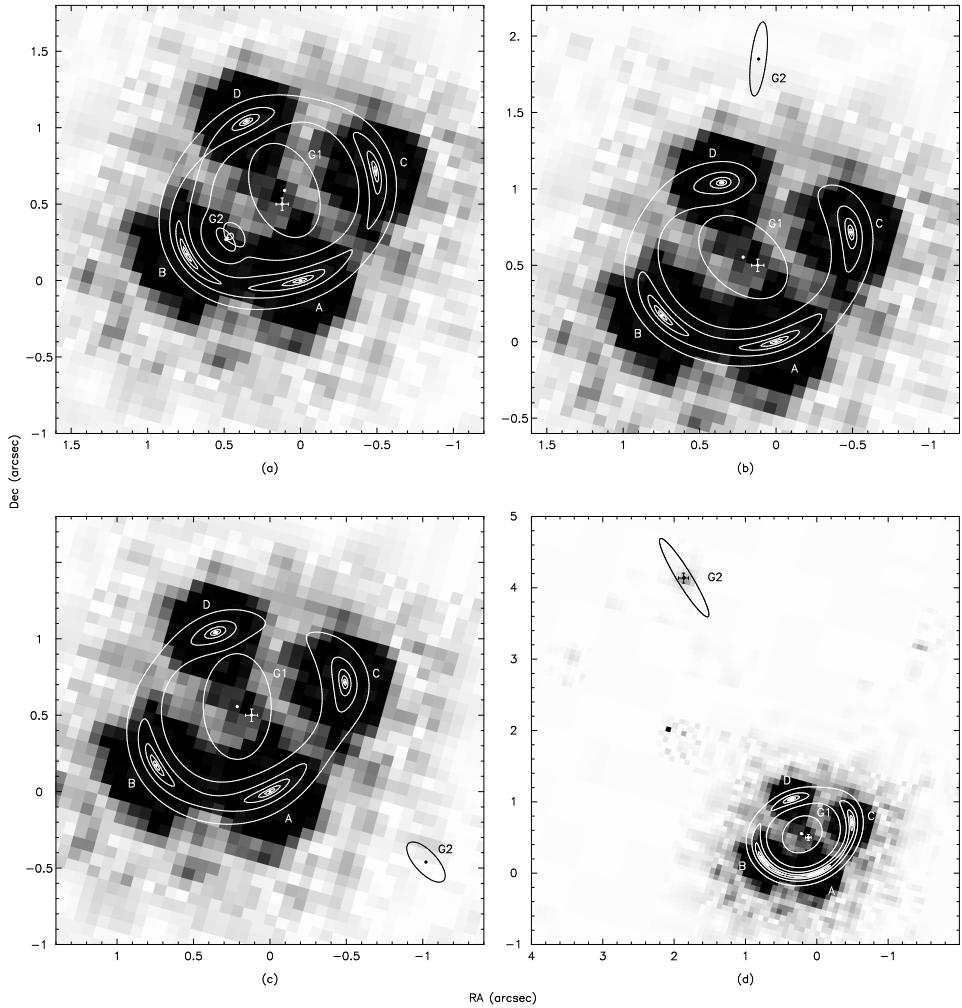}}
\caption[]{Image configuration of two-galaxy lens models (Table 4). 
The model image is overlaid on the HST NICMOS/F160W image. The five contour 
levels correspond to 40, 80, 160, 320, and 640 pc from the center of the QSO 
on the source plane. In the NICMOS image, light (probably from the primary 
lensing galaxy) is clearly present near the geometrical center of the 
Cloverleaf. The ellipses representing the galaxies correspond to 
a surface density of unity convergence ($\kappa=1$). 
The observed positions of the QSO components and the primary galaxy
are represented by dots with error bars (note that the error bars for the QSO
components are too small to be visible). The mass centers of the model 
galaxies are also represented by dots. North is on the top and
east is to the left.
(a) Model 1: The secondary galaxy (G2) is positioned southeast of the primary 
galaxy (G1). (b) Model 2: G2 is positioned north of G1. (c) Model 3: G2 is 
positioned southwest of G1. (d) Model 4: In this model a detected galaxy 
(object 14; Kneib et al.\ 1998a,b) is used as the secondary galaxy.
The observed position of the secondary galaxy is represented by a dot with 
error bars.}
\end{figure}

\begin{figure}
\centerline{\psfig{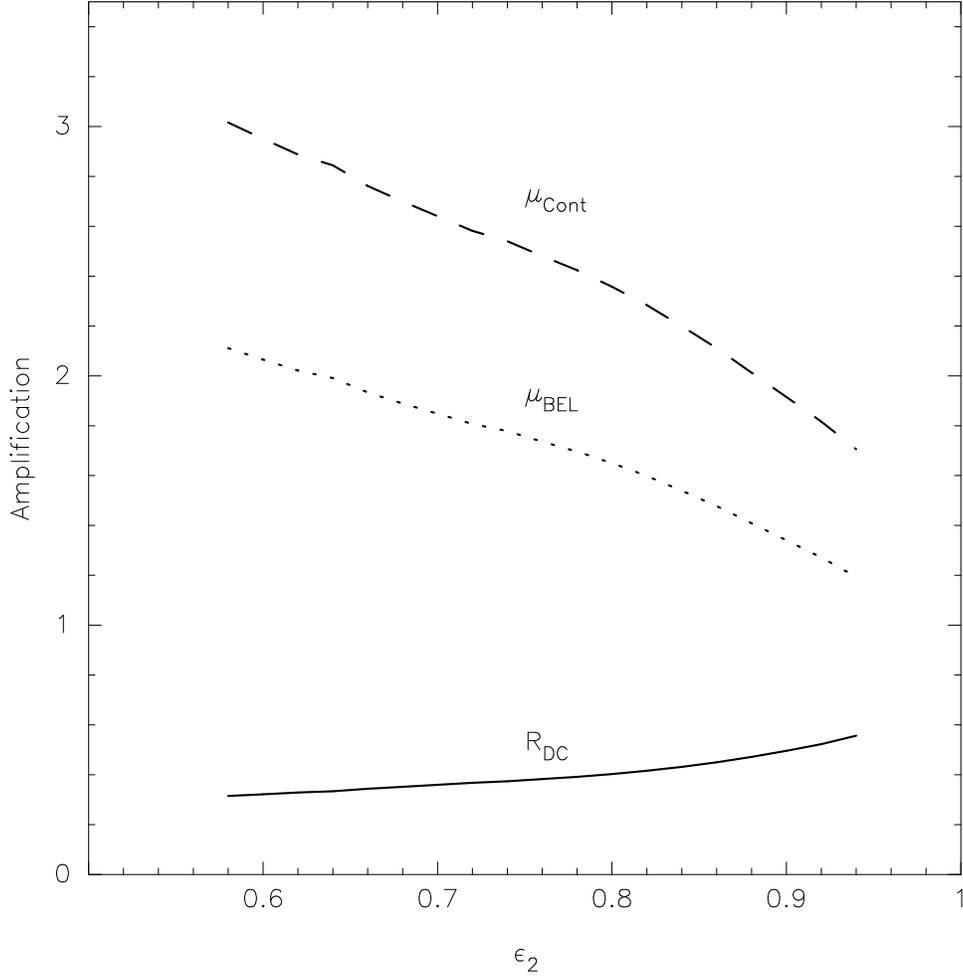}}
\caption[]{ Predicted relative magnification of component D ($R_{DC}$)
and the required microlensing magnification factors of the continuum and
the BELR as a function of $\epsilon_2$ (G2's ellipticity) in the two-galaxy
lens model of \S3.3.2. The model requires that the BELR as well as the 
continuum be amplified (with the continuum being more amplified) by 
microlensing in order to be consistent with the dust-extinction-corrected
relative amplification of T97 (see \S 3.3.2). A model with a smaller 
$\epsilon_2$ predicts a larger microlensing amplification.}
\end{figure}

\begin{figure}
\centerline{\psfig{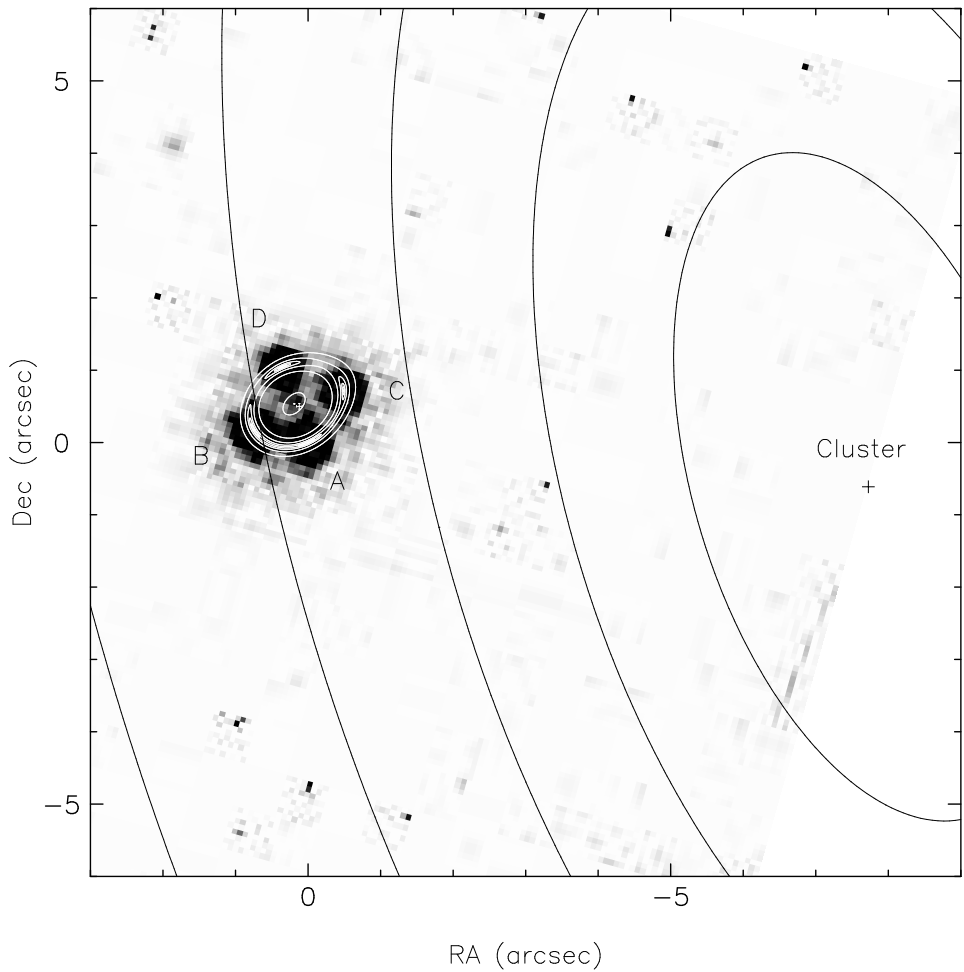}}
\caption[]{Galaxy+Cluster Model 1 (Table 5). The cluster's mass center
is at $8''$ (fixed) west of the lensing galaxy. The contours of the cluster
mass distribution correspond to convergence $\kappa = 0.4$, 0.5, 0.6, 0.7,
and 0.8 from east to west. The five contour levels of the image correspond to 
20, 40, 80, 160, and 320 pc from the center of the QSO on the source plane.
North is on the top and east is to the left.}
\end{figure}
\end{document}